\documentclass[conference]{IEEEtran}
\usepackage{cite}
\usepackage{amsmath,amssymb,amsfonts}
\usepackage{algorithmic}
\usepackage{graphicx}
\usepackage{textcomp}
\usepackage{xcolor}
\usepackage{float}
\usepackage[caption=false]{subfig}
\usepackage{booktabs}
\usepackage{tabulary}
\usepackage{array}
\usepackage{makecell}
\usepackage{tablefootnote}
\usepackage{graphics}

\usepackage{amssymb}
\usepackage{pifont}
\newcommand{\cmark}{\ding{51}}%
\newcommand{\xmark}{\ding{55}}%
\newcommand{\omark}{\ding{108}}%
\newcommand{\bigO}{\mathcal{O}}

\usepackage[flushleft]{threeparttable}

\usepackage[all]{background}
\usepackage{stackengine}
\setstackEOL{\\}
\setstackgap{L}{\normalbaselineskip}
\SetBgContents{\color{blue}{\tiny \Longstack{PREPRINT - accepted at 29th IFIP/IEEE International Conference on Very Large Scale Integration (VLSI-SoC 2021)}}}
\SetBgPosition{3.5cm,1cm}
\SetBgOpacity{1.0}
\SetBgAngle{0}
\SetBgScale{1.8}

\usepackage{mathtools}

\newcolumntype{L}{>{\centering\arraybackslash}m{1.35cm}}
\def\BibTeX{{\rm B\kern-.05em{\sc i\kern-.025em b}\kern-.08em
		T\kern-.1667em\lower.7ex\hbox{E}\kern-.125emX}}

\begin{document}
	\bstctlcite{IEEEexample:BSTcontrol}
	
	
	\title{Logic Locking at the Frontiers of Machine Learning: A Survey on Developments and Opportunities\vspace{-0.2in}}
	
	\author{\IEEEauthorblockN{Dominik~Sisejkovic, Lennart M. Reimann, Elmira Moussavi, Farhad~Merchant,~and~Rainer~Leupers}
		\IEEEauthorblockA{\textit{Institute for Communication Technologies and Embedded Systems, RWTH Aachen University, Germany} \\
			\{sisejkovic, reimannl, moussavi, merchantf, leupers\}@ice.rwth-aachen.de}\vspace{-0.3in}
	}
	
	\maketitle
	
	\begin{abstract}
		In the past decade, a lot of progress has been made in the design and evaluation of
		logic locking; a premier technique to safeguard the integrity of integrated circuits throughout the electronics supply chain. 
		However, the widespread proliferation of machine learning has recently introduced a new pathway to evaluating logic locking schemes. This paper summarizes the recent developments in logic locking attacks and countermeasures at the frontiers of contemporary machine learning models. Based on the presented work, the key takeaways, opportunities, and challenges are highlighted to offer recommendations for the design of next-generation logic locking.
	\end{abstract}
	
	\begin{IEEEkeywords}
		logic locking, machine learning, neural networks, reverse engineering, genetic algorithms
	\end{IEEEkeywords}
	
	
	\section{Introduction}
	The involvement of untrusted parties in the modern Integrated Circuit (IC) supply has given rise to a plethora of security concerns, including Intellectual Property (IP) piracy, reverse engineering, counterfeiting, and hardware Trojans~\cite{rostami2014primer,hardwareTrojansLessonsLearned2016}. Consequently, a variety of countermeasures have been introduced, including IC metering~\cite{metering2012}, split manufacturing~\cite{splitManufacturing2013}, camouflaging~\cite{camouflaging2013}, and logic locking~\cite{evolutionOfLogicLocking2017}. Among these, only logic locking can protect a design against all untrusted parties in the supply chain~\cite{keynoteLogicLocking2020,evolutionOfLogicLocking2017}.
	
	\subsection{Logic Locking: A Brief Overview}
	Logic locking performs design manipulations by binding the correct functionality of a hardware design to a secret key that is only known to the legitimate IP owner. Hereby, both the original functionality and the structure of the design remain concealed while passing through the hands of external design houses and the foundry. In the past decade, various security aspects of logic locking have been thoroughly evaluated through the introduction of key-recovery attacks~\cite{threatsDecadeLater2019,removalAttacks2020}, among which the Boolean satisfiability (SAT) attack has gained a lot of attention~\cite{originalSatAttack2015}. This has led to a division of logic locking into \textit{pre- and post-SAT schemes}. Pre-SAT schemes were focusing on specific security features, such as random XOR/XNOR key-gate insertion~\cite{epic2008}, thwarting the path-sensitization attack~\cite{sll2} or maximizing output corruption for incorrect keys~\cite{faultAnalysisBasedLL2015}. With the introduction of SAT-based attacks, the design objective has shifted towards achieving SAT-resilience, resulting in a new generation of schemes, including SARLock~\cite{sarLock2016}, Anti-SAT~\cite{antiSat2019}, CASLock~\cite{caslock2019}, SFLL~\cite{sfll2017}, and others~\cite{threatsDecadeLater2019}. 

	\subsection{The Advent of Machine Learning}
	With the advent of efficient and easy-to-use Machine Learning (ML) models, ML-based techniques have been gradually introduced into various hardware-security domains~\cite{surveyMLforHwTrojanDetection2020,MlforHWSec2018}. The latest efforts in the logic locking community have been invested in challenging the security properties of locking schemes using ML. Recent works were able to efficiently attack pre- and post-SAT schemes~\cite{gnnunlock2020, sisejkovic2020challenging, SAIL2019, genUnlock2019,azar2020nngsat,particleSwarmBasedLL2020,BOCANet2019,SURF2019,icDeobfuscationRuntimeEstimationWithDeepLearning2020}. {The introduction of ML-based tools for the security analysis of logic locking \textit{has opened up a new chapter} in the design of locking schemes and attacks, thereby initiating the start of the \textit{post-ML} locking-scheme era. Herewith, the ML ecosystem offers a novel path to uncover hidden vulnerabilities and provide new directions in the development of future ML-resilient locking schemes.
		
		\textbf{Contributions}
		The ML era has undoubtedly initiated a new stage in logic locking design and evaluation. In this paper, we review all major developments in the domain of \textit{ML-based attacks and countermeasures in logic locking}, and analyze major challenges and research opportunities. Note that a comprehensive overview of the state of pre-ML schemes and attacks can be found in~\cite{evolutionOfLogicLocking2017, keynoteLogicLocking2020, removalAttacks2020, threatsDecadeLater2019, systematicOverviewLogicLocking2019}.

		The rest of this paper is organized as follows. Section~\ref{sec:background} introduces the relevant background on logic locking. Section~\ref{sec:ll-in-ml-era} reviews the major developments in ML-based logic locking attacks and compiles a summary of the open challenges and opportunities. Finally, Section~\ref{sec:conclusion} concludes the paper.

		
		\section{Background}\label{sec:background}
		This section introduces the preliminaries on classification, working principles and attack models of logic locking.
		
		\subsection{Classification}
		Logic locking can be generally classified into two orthogonal classes: \textit{combinational} and \textit{sequential}~\cite{systematicOverviewLogicLocking2019}. Combinational logic locking performs key-dependent manipulations in the combinational path of a design. On the other hand, sequential logic locking focuses on transforming and obfuscating the state space of a circuit. As the reviewed work operates in the domain of combinational locking, in the rest of this work, the term \textit{logic locking} refers to combinational locking schemes.
		
		\subsection{Working Principles}
		The idea of logic locking lies in the functional and structural manipulation of a hardware design that creates a dependency to an activation key, hereby trading area, power, and delay for security. If the correct key is provided, the locked design will perform as originally intended \textit{for all input patterns}. Otherwise, an incorrect key will yield an incorrect output for at least \textit{some input patterns}. Logic locking can be performed on different design levels. However, typically logic locking is deployed on a gate-level netlist through the insertion of additional gates (known as \textit{key gates}) or more complex structures. A visual example of a locked design is shown in Fig.~\ref{fig:enc-example}~(b). Here, the original netlist in Fig.~\ref{fig:enc-example}~(a) is locked through the insertion of two key-controlled gates in the form of an XOR and XNOR (XOR + INV) gate, marked as $KG_{1}$ and $KG_{2}$, respectively. To understand the functional implications of the key gates, let us consider the gate $KG_{1}$. This key gate takes two inputs: the original wire $x$ (output of gate $G_{1}$) and the input key bit $k_{1}$. If a correct key value is set, i.e., $k_{1}=0$, the value of $x$ is preserved and forwarded to $x'$. However, if an incorrect key value is set, i.e., $k_{1}=1$, the value of $x$ is inverted, leading to incorrect output values. Based on this concept, throughout the past decade, a variety of locking schemes have been introduced, based on XOR, XNOR, AND, OR, and MUX gates as well as more elaborate structures~\cite{evolutionOfLogicLocking2017}.

		\begin{figure}[t]
			\centering
			\subfloat[Original Circuit]{
				\includegraphics[width=0.35\columnwidth]{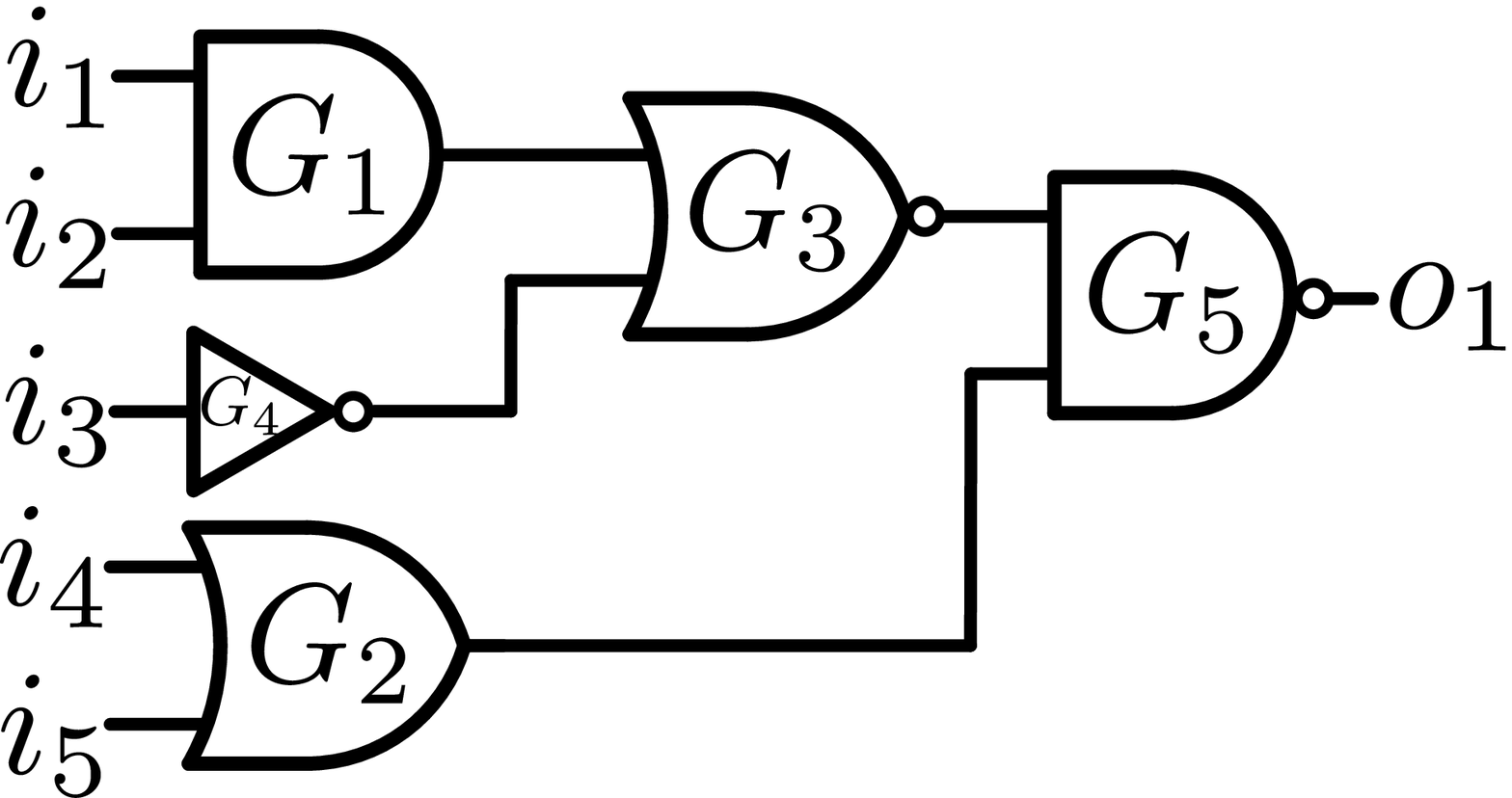}
			}
			\subfloat[Locked Circuit]{
				\includegraphics[width=0.4\columnwidth]{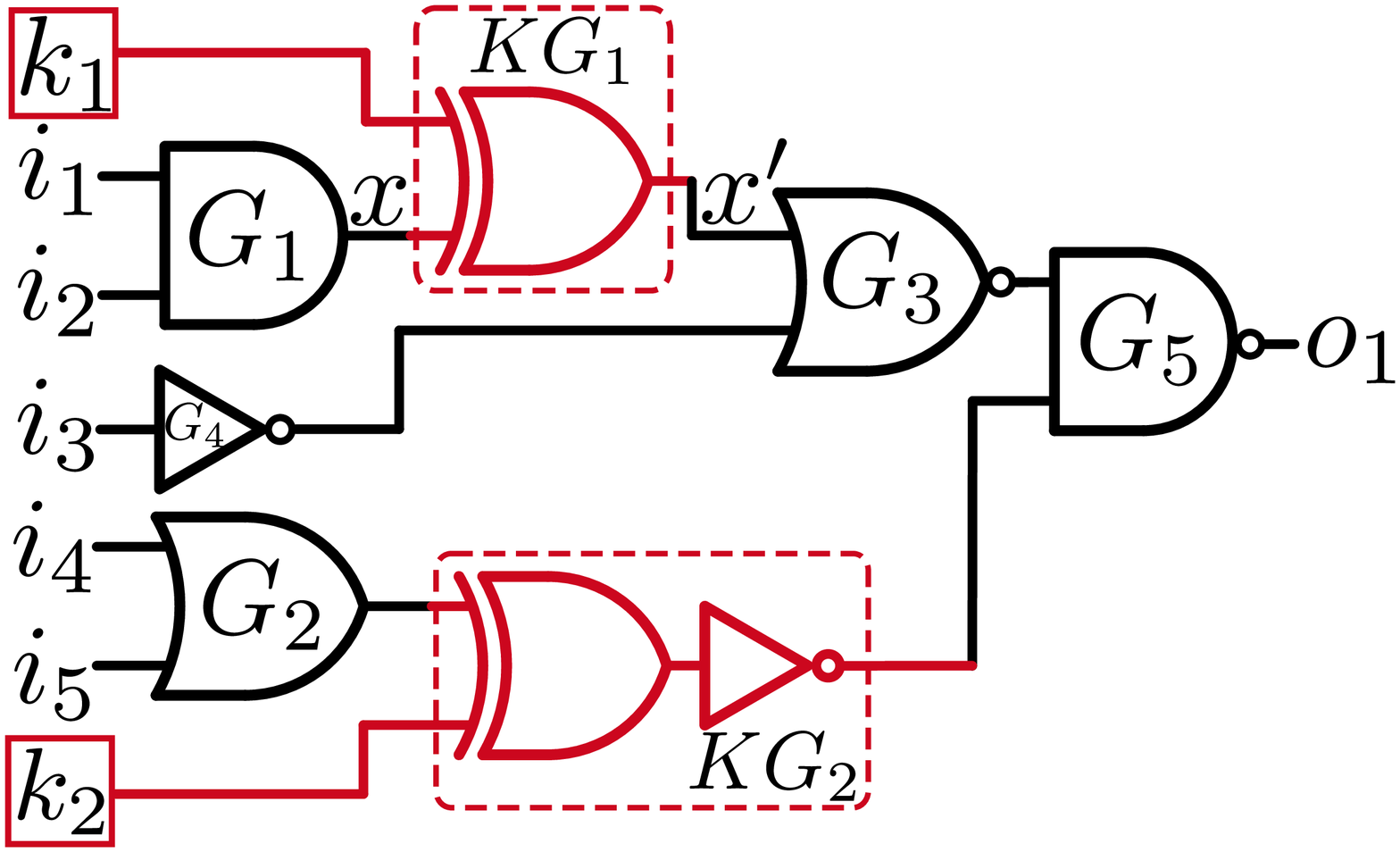}
			}
			\caption{Example: logic locking using XOR/XNOR gates.}
			\label{fig:enc-example}
			\vspace{-0.1in}
		\end{figure}
		
		

		\subsection{Logic Locking in the IC Supply Chain}
		The role of logic locking in the IC supply chain is demonstrated in Fig.~\ref{fig:supplychain}. Based on a trusted Register-Transfer Level (RTL) design, the legitimate IP owner performs logic synthesis to generate a gate-level netlist. At this point, logic locking is deployed, resulting in a locked netlist and an activation key. Typically, after the netlist is locked, another synthesis round is performed to facilitate the structural integration of scheme-induced netlist changes. Therefore, we can differentiate between the pre-resynthesis and post-resynthesis netlist. The former is locked but not resynthesized, while the latter is locked and resynthesized. In the next step, the locked netlist proceeds in the untrusted part of the supply chain. This often includes an untrusted external design house (for layout synthesis) and the foundry.  After fabrication, the produced IC is returned to the IP owner for activation. Herewith, logic locking protects a design by concealing its functional and structural secrets in the activation key, thereby bridging the untrusted regime gap. 
		In terms of hardware Trojans, it is assumed that a sound understanding of the design's functionality and structure is required to insert an intelligible, controllable and design-specific Trojan (e.g, a targeted denial-of-service attack). However, functionality-independent Trojans remain viable. These include, e.g., the manipulation of the circuit's physical characteristics, leading to performance or reliability degradation. In the former case, finding the activation key is a prerequisite for successfully performing the reverse-engineering process. 

		\begin{figure}[!t]
			\centering
			\includegraphics[width=\columnwidth]{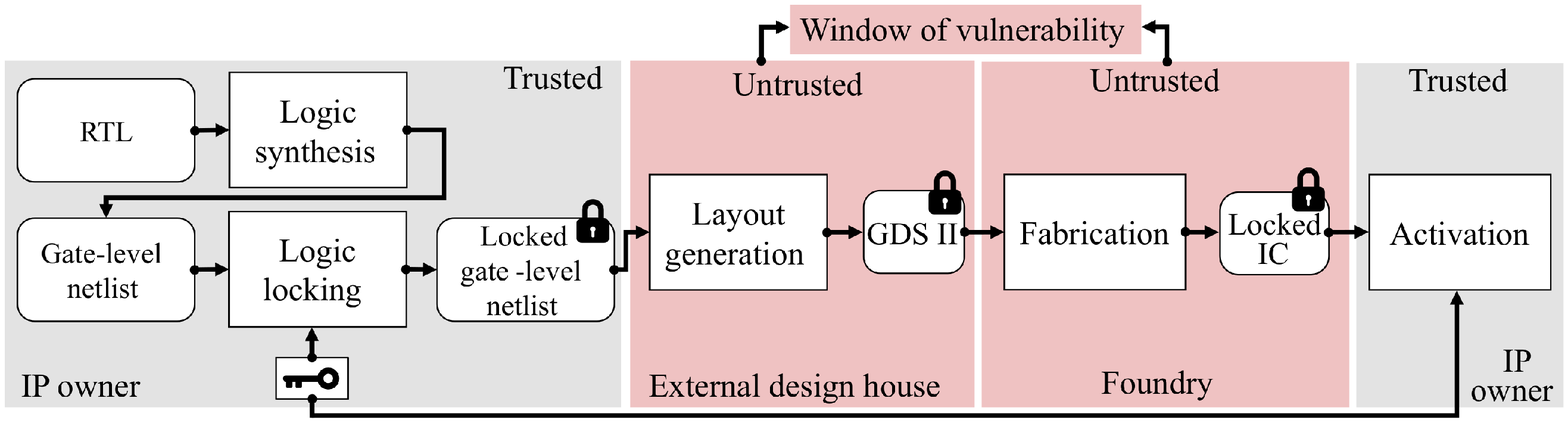}
			\caption{Logic locking in the IC design and fabrication flow.}
			\label{fig:supplychain}
			\vspace{-0.15in}
		\end{figure}
		\begin{figure*}[t]
			\centering
			\includegraphics[width=\textwidth]{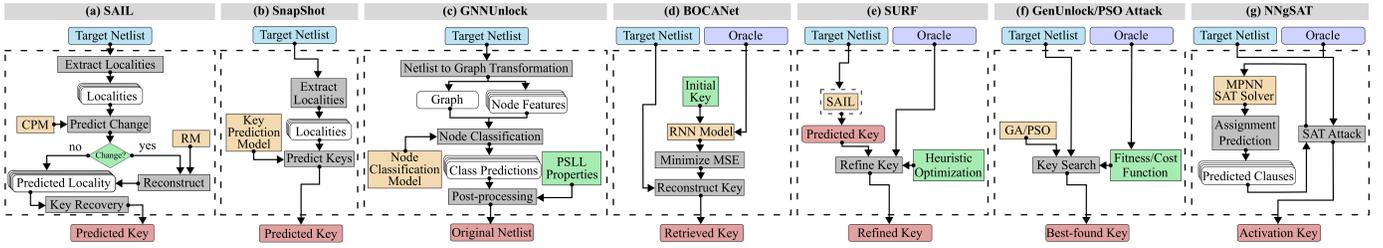}
			\caption{ML-based attacks on logic locking (deployment phase).}
			\label{fig:attack-visualization}
			\vspace{-0.1in}
		\end{figure*}
		
		\subsection{Attack Model}
		The attack model includes the following: ($i$) the attacker has access to the locked netlist, either as an untrusted design house or by reverse engineering the layout, ($ii$) the location of the key inputs (pins) is known, ($iii$) the deployed locking scheme is known, and ($iv$) the attacker has access to an activated IC to use as oracle for retrieving golden Input/Output (I/O) patterns.
		
		%
		The fourth assumption is the differentiating factor that classifies all key-retrieval attacks into \textit{oracle-less} and \textit{oracle-guided} attacks. Oracle-less attacks assume that an activated design is not available. This is often the case in low-volume production for security-critical applications~\cite{force2005high}. On the contrary, the oracle-guided scenario assumes the availability of an activated IC, thereby representing a high-volume production setting~\cite{pilato2020assure}.
		Moreover, sometimes, an attack assumes the availability of only a few golden I/O patterns, e.g., in the form of test vectors. Due to the available knowledge that is provided by these patterns, these attacks fall into the oracle-guided class. As discussed in the next section, ML-based attacks have been explored in both classes. Henceforth, we refer to the gate-level netlist under attack as the \textit{target} netlist.
		
		
		
		

		%
		
		\section{Logic Locking in the Machine Learning Era}\label{sec:ll-in-ml-era}
		This section describes the recent developments of ML-based applications in the logic locking domain. 
		
		
		
		\subsection{ML-Based Attacks}
		Previous work in this domain has mostly been focusing on the development of novel ML-based attacks on logic locking, both for the oracle-less and oracle-guided model. A simplified visualization of all reviewed attack flows is presented in Fig.~\ref{fig:attack-visualization}. Hereby, the attacks are presented in the \textit{deployment} stage (after training). An exhaustive comparison of the attacks is summarized in Table~\ref{table:attacks}. The comparison lists the attacks in order of appearance per attack class. For convenience, a glossary of the used acronyms is given in Table~\ref{table:glossary}. 
		
		The following review reflects the descriptions in Fig.~\ref{fig:attack-visualization} and Table~\ref{table:attacks}, thereby only focusing on the \textit{basic attack mechanisms}. All other details can be found in the mentioned \textit{comparison table} and the provided references. 
		
		\subsubsection{Oracle-Less Attacks}
		ML-based attacks in this class have exploited \textit{scheme-related structural residue} to identify a correct key-bit value or the locking circuitry itself. This category includes SAIL~\cite{SAIL2019}, SnapShot~\cite{sisejkovic2020challenging}, and GNNUnlock~\cite{gnnunlock2020}.
		
		\textbf{SAIL:} This attack deploys ML algorithms to retrieve local logic structures from the post-resynthesis target netlist to predict the correct key values (Fig.~\ref{fig:attack-visualization}~(a)). The attack targets XOR/XNOR-based logic locking exclusively. Hereby, the attack exploits two leakage points of the locking flow: ($i$) the deterministic structural changes induced by logic synthesis around the key-affected netlist gates and ($ii$) the nature of XOR/XNOR-based locking (XOR for key bit 0 and XNOR for key bit 1). Therefore, the attack encloses two components: the Change Prediction Model (CPM) and the Reconstruction Model (RM). For each netlist subgraph around a key input (known as \textit{locality}), CPM is trained to predict whether a synthesis-induced change has occurred. If a change is predicted, RM is deployed to reconstruct the pre-resynthesis netlist structures, i.e., it reverses the effect of the synthesis process. Finally, based on the intrinsic nature of XOR/XNOR-based locking, SAIL extracts the correct key value.
		
		\textbf{SnapShot:} The attack utilizes a neuroevolutionary approach to automatically design suitable neural networks to \textit{directly} predict correct key values from a locked post-resynthesis target netlist (Fig.~\ref{fig:attack-visualization}~(b)). The attack exploits the structural alterations induced by locking schemes to learn a correlation between the key-induced structural patterns and the key value. Compared to SAIL, SnapShot implements an end-to-end ML approach, thereby having the advantage of being applicable to any locking scheme as well as not relying on learning specific transformation rules of the logic synthesis process.
		
		\textbf{GNNUnlock:} The attack leverages a graph neural network to identify all the gates in a post-resynthesis target netlist that belong to the locking circuitry (Fig.~\ref{fig:attack-visualization}~(c)). Therefore, compared to SAIL and SnapShot, GNNUnlock learns to differentiate locking gates from regular gates instead of learning a correlation between the netlist sub-graphs and the correct key. To enhance the removal accuracy, after identification, a deterministic post-processing mechanism is deployed to remove the gates depending on the intrinsic features of the underlying schemes. Hereby, GNNUnlock has specifically been trained and deployed to target SAT-attack-resilient schemes, i.e., Provably Secure Logic Locking (PSLL). Herein lies the success of the attack: PSLLs often induce isolable structural netlist changes to produce a specific SAT-resilient circuit behavior. In the pre-ML era, it has been long assumed that PSLLs can be protected through additional locking and resynthesis.
		\begin{table}[b]\footnotesize
			\centering
			\vspace{-0.2in}
			\caption{Glossary}
			\label{table:glossary}
			\resizebox{\columnwidth}{!}{%
				\begin{tabular}
					{l|l||l|l}
					\toprule
					\textbf{Acronym} & \textbf{Definition}&\textbf{Acronym} & \textbf{Definition}\\
					\midrule
					\midrule
					AND-OR & AND/OR-based LL~\cite{iolts2014} & LUT & Lookup table \\
					Anti-SAT & Anti-Boolean satisfiability~\cite{antiSat2019} & MLP & Multi-layer perceptron\\
					C & Combinational circuits & MPNN & Message-passing neural network\\
					CNN & Convolutional neural network  & OG & Oracle-guided attack\\
					CS & Logic-cone-size-based LL~\cite{coneBasedLocking2018} & OL & Oracle-less attack\\
					D-RNN & Deep recurrent neural network & OLL & Optimal LL~\cite{optimalLL2018}\\
					FLL & Fault analysis-based LL~\cite{faultAnalysisBasedLL2015} & PSO & Particle swarm optimization\\
					FU & Functional &  RLL & Random LL~\cite{epic2008}\\
					GA & Genetic algorithm  & S & Sequential circuits\\
					GL & Gate level & SFLL-HD & Stripped functionality LL~\cite{sfll2017}\\
					GNN & Graph neural network & SLL & Strong (secure) LL~\cite{sll2}\\
					LL & Logic locking  & ST & Structural\\
					LSTM & Long short-term memory  &  TTLock & Tenacious and traceless LL~\cite{TTLock2017}\\
					%
					\bottomrule
					\bottomrule
				\end{tabular}
			}
		\end{table}
		\begin{table*}[hbtp]\footnotesize
			\setlength{\tabcolsep}{2pt}
			\caption{Overview of ML-Based Attacks on Logic Locking}
			\label{table:attacks}
			\resizebox{\textwidth}{!}{%
				\begin{threeparttable}
					\centering
					\begin{tabular}{l||>{\centering\arraybackslash}m{1.3cm}>{\centering\arraybackslash}m{0.5cm}>{\centering\arraybackslash}m{1.7cm}>{\centering\arraybackslash}m{1.2cm}>{\centering}m{1.7cm}>{\centering\arraybackslash}m{1.2cm}>{\centering}m{1cm}>{\centering}m{1.2cm}>{\centering\arraybackslash}m{2.1cm}>{\centering\arraybackslash}m{1.7cm}>{\centering\arraybackslash}m{1.5cm}>{\centering\arraybackslash}m{1.7cm}}
						\toprule
						\textbf{Attack}& \textbf{Objective}&\textbf{Class}& \textbf{Level/IC Type/Attack Basis}&\textbf{ML Model}&\textbf{Benchmarks}\textsuperscript{16}&\textbf{Evaluated Schemes}&\textbf{Scheme Independent}&\textbf{Exact Output}\textsuperscript{15}&\textbf{Evaluate Key Length in Bits}&\textbf{\% Accuracy [min, max]}&\textbf{Time Complexity\textsuperscript{17}}&\textbf{Known Protection}\\
						\midrule
						
						SAIL~\cite{SAIL2019}& key retrieval & OL&GL/C,S\textsuperscript{1}/ST&Random Forest& ISCAS'85&RLL, SLL, CS& \xmark&\xmark&$\{4,\cdots,192\}$\textsuperscript{4}&$[66.89, 73.88]$&$\bigO(l)$\textsuperscript{13}& UNSAIL~\cite{unsail2020}, SARO~\cite{alaql2020scalable}\textsuperscript{22}\\\hline
						
						SnapShot~\cite{sisejkovic2020challenging}& ($i$) GSS and ($ii$) SRS key retrieval\textsuperscript{14}& OL&GL/C,S/ST&MLP, CNN, GA& ISCAS'85, ITC'99, Ariane RV&RLL& \cmark&\xmark&$\{64\}$&$[57.71, 61.56]^{(i)}$, $[71.57, 81.67]^{(ii)}$&$\bigO(l)$\textsuperscript{13}& D-MUX~\cite{sisejkovic2021deceptive}\\\hline
						
						GNNUnlock~\cite{gnnunlock2020}&key gates removal&OL&GL/C,S/ST&GNN&ISCAS'85, ITC'99&Anti-SAT, TTLock, SFLL-HD&\xmark\textsuperscript{12}&\xmark&$\{8,\cdots,128\}$&$[100.00]$\textsuperscript{11}&$\bigO(n)$\textsuperscript{13}&\omark\\
						\midrule
						\midrule
						
						BOCANet~\cite{BOCANet2019, BOCANet2019_2}& key retrieval & OG\textsuperscript{2}&GL/C/FU&D-RNN \& LSTM& Trust-Hub, ISCAS'85&RLL&\cmark&\xmark & $\{32,64,128,256\}$&$[89.00, 100.00]$&$\bigO(\alpha)$\textsuperscript{19}&\omark\\\hline
						
						SURF~\cite{SURF2019}&key retrieval& OG &GL/C,S\textsuperscript{1}/ST,FU&SAIL \& heuristic optimization&ISCAS'85&RLL, SLL, CS&\xmark\textsuperscript{3}&\xmark&$\{4,\cdots,192\}$\textsuperscript{4} &$[90.58,98.83]$&$\bigO(t)$\textsuperscript{13}&UNSAIL~\cite{unsail2020}\\\hline
						
						GenUnlock~\cite{genUnlock2019}&key retrieval& OG &GL/C,S/FU&GA&ISCAS'85, MCNC&RLL,SLL, AND-OR, FLL&\cmark&\xmark&$\{8,\cdots, 1618\}$\textsuperscript{7}&$[100.00]$\textsuperscript{8}&$\bigO(\beta)$\textsuperscript{20}&\omark\\\hline
						
						NNgSAT~\cite{azar2020nngsat}&key retrieval& OG &GL/C,S/FU&MPNN&ISCAS'85, ITC'99&SAT-hard\textsuperscript{5}&\cmark&\cmark&n/a\textsuperscript{5}&$[93.50]$\textsuperscript{6}&$\bigO(\lambda)$\textsuperscript{18}&\omark\\	\hline
						
						PSO Attack~\cite{particleSwarmBasedLL2020}&key retrieval& OG\textsuperscript{9} &GL/C,S/FU&PSO&ISCAS'85, ITC'99&RLL, OLL&\cmark&\xmark&$\{64,128\}$&$[82.07,99.80]$\textsuperscript{10}&$\bigO(\gamma)$\textsuperscript{21}&point-function locking~\cite{antiSat2019,sfll2017}\\
						
						\bottomrule
					\end{tabular}
					\begin{tablenotes}
						\scriptsize
						\item [1]~~The attack is in theory applicable to sequential circuits, however no evaluation has been performed yet.
						\item [2]~~The attack relies on having access to at least some golden input/output patterns ($<0.5\%$ of total I/O pairs).
						\item [3]~~In theory, the key-refinement search algorithm could be utilized based on any seed key. However, this has not been addresses thus far.
						\item [4]~~The following key lengths have been evaluated: $\{4,8,16,24,32,64,96,128,192\}$.
						\item [5]~~$n\times m$ bitwise multipliers $(8<m,n<32)$,  $n\times m$ crossbar network of 2-to-1 MUXes $(16<m, n<36)$,  $n$-input LUTs built by 2-to-1 MUXes $(n<16)$, and $n$-to-1 AND-trees.
						\item [6]~~Indicates the percentage of successfully de-obfuscated circuits compared to the baseline~\cite{originalSatAttack2015}.
						\item [7]~~The key length is selected based on an area overhead of 5\%, 10\% or 25\%.
						\item [8]~~The quality of the retrieved approximate keys is quantified by a user-defined output-fidelity measure.
						\item [9]~~The attack relies on an oracle without access to the scan chain.
						\item [10]~Accuracy refers to the average number of cases where the retrieved key results in 0\% erroneous outputs for $10^{6}$ random patterns.
						\item [11]~The accuracy refers to the successful removal of the locking circuitry (not the key retrieval).
						\item [12]~The attack has not been evaluated for other locking schemes so far and the post-processing steps are scheme-specific.
						\item [13]~Notation: the key length $l$, the number of netlist nodes $n$, the total number of iterations $t=p\cdot{n} + p\cdot{w}+r\cdot{l}\cdot{i}\cdot{n}+r\cdot{l}\cdot{i}\cdot{p}$, where $p$ is the number of output pins, $i$ is the number of IO pairs, $r$ is the number of runs, and $w$ is the number of wires.
						\item [14]~GSS refers to the generalized set scenario which trains the ML model based on a set of locked benchmarks that are different from the target. SRS captures the self-referencing scenario where the training data is generated by re-locking the target benchmark.
						\item [15]~If the output of the attack is an exact result, the attack can guarantee a 100\% correct deobfuscation for the complete I/O space.
						\item [16]~ISCAS'85~\cite{iscas2}, MCNC~\cite{mcnc}, ITC'99~\cite{itc99}, RISC-V Ariane core~\cite{ariane2019}, and Trust-Hub~\cite{TrustHub1, TrustHub2}.
						\item [17]~If the time complexity can be clearly determined, it refers to the time complexity after the training process.
						\item [18]~The execution time of the SAT attack is $\sum_{i=1}^{\lambda}{t_{i}}$, where $\lambda$ is the number of iterations and $t_{i}$ the time required for one SAT-solver call. $\lambda$ depends on the characteristics of the search space and the branching preferences of the SAT solver. Therefore, the time complexity of the attack is typically measured in terms of $\lambda$. More details can be found in~\cite{satAttackTimeComplexity2020}.
						\item [19]~The complexity is linear to the number of training samples $\alpha$, as the final key is determined based on the MSE of the trained outputs and the key-induced generated outputs.
						\item [20]~$\beta=g\cdot{p}\cdot{l}$, where $g$ is the number of generations, $p$ is the population size, and $l$ is the key length. Note that the complexity changes with any adaptations of the GA.
						\item [21]~$\gamma=g\cdot{p}\cdot{\delta}$, where $g$ is the number of generations, $p$ is the population size, and $\delta$ the complexity of performing circuit simulation for the fitness evaluation.
						\item [22]~An empirical evaluation of the resilience against SAIL has not been presented in the paper; only a discussion based on a proposed metric system has been provided.
					\end{tablenotes}
				\end{threeparttable}
			}
			\vspace{-0.1in}
		\end{table*}%
		
		\subsubsection{Oracle-Guided Attacks} Due to the availability of an oracle, existing ML-based attacks in this class have mostly exploited functional features of the target IC. This class includes the following attacks: BOCANet~\cite{BOCANet2019, BOCANet2019_2}, SURF~\cite{SURF2019}, GenUnlock~\cite{genUnlock2019}, NNgSAT~\cite{azar2020nngsat}, and the PSO-guided attack~\cite{particleSwarmBasedLL2020}.
		
		\textbf{BOCANet:}~This attack leverages Recurrent Neural Networks (RNN) based on long short-term memory to construct a correct activation key (Fig.~\ref{fig:attack-visualization}~(d)). The ML model is trained on a sequence of I/O observations  taken from an activated IC, thereby learning the functional I/O mapping of the circuit, i.e., its Boolean function. Once trained, the key retrieval consists of two steps. First, a random key is applied to the model as input. Second, the initial key value is subsequently updated based on the Mean-Squared-Error (MSE) of the trained outputs and the newly generated outputs that are affected by the introduced key. Note that the ML model can be utilized to predict correct inputs or outputs as well. BOCANet exploits the functional effect a correct key has on generating a correct I/O mapping.
		
		\textbf{SURF:}~This attack is based on a joint structural and functional analysis of the circuit to retrieve the activation key (Fig.~\ref{fig:attack-visualization}~(e)). The ML-aspect of the attack lies in \textit{utilizing the SAIL attack} to generate the pre-resynthesis netlist structures and a seed key. Afterwards, based on the outputs of SAIL, SURF iteratively refines the key by means of a structure-aware greedy optimization algorithm guided by a functional simulation of the obfuscated netlist and a set of golden I/O pairs. The optimization is guided by the observation that specific key gates only affect a specific set of outputs. Thus, the key bits can be partitioned based on which output they affect. Performing a systematic perturbation of the key bits can lead to a more refined key. The success of the heuristic is grounded in the limited local effects that traditional locking schemes have on the value of the output for incorrect keys.
		
		\textbf{GenUnlock:}~The attack flow of GenUnlock leverages a Genetic Algorithm (GA)-based exploration of suitable activation keys for a locked circuit (Fig.~\ref{fig:attack-visualization}~(f)). The heuristic search is steered by the key fitness that is computed based on the matching ratio of the key on the golden I/O training set. Through multiple generations, the fitness of the key population is subsequently improved through the application of genetic operators (selection, crossover, and mutation). Once the accepted tolerance for the correctness of the key is reached, the algorithm returns the set of the fittest keys. Similarly to SURF, GenUnlock exploits the fact that a heuristic key-refinement procedure eventually leads to more accurate activation keys. 
		
		\textbf{NNgSAT:}~The main objective of NNgSAT is the deployment of a Message-Passing Neural Network (MPNN) to facilitate the resolution of SAT-hard circuit structures during the application of a SAT-based attack (Fig.~\ref{fig:attack-visualization}~(g)). The motivation is driven by the fact that common SAT solvers run into scalability issues when tackling hard-to-be-solved locked circuit structures, e.g, multipliers and AND trees. Therefore, in NNgSAT, a neural network is trained to predict the satisfying assignment on a set of SAT-hard cases. In deployment, the SAT-attack flow offloads the SAT-hard problems to the trained model to speed up the attack procedure. The effectiveness of NNgSAT lies in the fact that it is possible to transfer prediction knowledge from learned clauses to unseen problems.  
		
		\textbf{PSO-guided Attack:}~This attack is based on a Particle Swarm Optimization (PSO) heuristic that searches through the key space directed by a selected cost function (Fig.~\ref{fig:attack-visualization}~(f)). The cost function is modeled as the Hamming distance between the golden and the obtained output responses (for a selected key). Therefore, the search algorithm relies on having access to an activated IC to compare against. A major motivator for this attack is its applicability without having access to an open scan chain, as this is often a limiting factor for SAT-based attacks. In essence, the PSO-guided attack and GenUnlock are similar in nature, as both rely on black-box evolutionary procedures guided by a functionality-driven objective function.
		%
		
		
		\subsection{ML-Resilient Schemes}
		\textbf{UNSAIL:}~This logic locking scheme has been developed to thwart attacks that target the resolution of structural transformations of logic synthesis~\cite{unsail2020}. The core idea of UNSAIL is to generate confusing training data that leads to false predictions in the CPM and RM modules of SAIL. This is realized through the additional manipulation of the netlist after synthesis to force the existence of \textit{equivalent} netlist sub-graph observations that are linked to \textit{different} key values.
		
		\textbf{SARO:}~The Scalable Attack-Resistant Obfuscation (SARO) operates in two steps~\cite{alaql2020scalable}. First, SARO splits the design into smaller partitions to maximize the structural alterations in the netlist. Second, a systematic truth table transformation is deployed to lock the partitions. In order to increase the complexity of pattern recognition attacks (such as SAIL), the transformations aim to maximize randomness in the netlist.
		
		\textbf{Point-Functions and PSO:} As mentioned in Table~\ref{table:attacks}, the PSO-guided attack is not applicable to point-function-based locking schemes. The reason is that this type of locking yields SAT-resilient behavior in which any incorrect key corrupts only \textit{a very limited amount of outputs}. Consequently, this behavior offers no advantage in the guidance of the heuristic search, as it does not yield differentiating fitness values. Note that the applicability of point functions depends on the design of the fitness function that is used to guide the heuristic.

		\textbf{D-MUX:} The recently introduced Deceptive Multiplexer (D-MUX) LL scheme builds on the concept of multiple MUX-based insertion strategies that create structural paths that are equally likely to be driven by 0 or 1 key values~\cite{sisejkovic2021deceptive}. Hence, D-MUX offers efficient protection against data-driven attacks.
		
		\subsection{Other Applications of ML}
		\textbf{Deobfuscation Runtime Prediction:}~Apart from ML-based attacks, machine learning has also found its way into other aspects of logic locking. A recent work has designed a framework named ICNet for the prediction of the key-retrieval runtime for SAT-based attacks using graph neural networks~\cite{icDeobfuscationRuntimeEstimationWithDeepLearning2020}. The framework obtains the predicted deobfuscation runtime based on the characteristics of the circuit topology. ICNet offers an end-to-end approach to evaluate the hardness of logic locking with respect to the SAT attack, thereby increasing the development efficiency of novel locking schemes. 
		
		\textbf{ML-Attack on Analog IC Locking:}~ML has started to have an impact on locking mechanisms even beyond digital circuits. The authors in~\cite{acharya2020attackOfTheGenes} have developed an oracle-guided attack on locked analog ICs using genetic algorithms. The approach has successfully broken all known analog logic locking techniques.
		
		\subsection{Lessons Learned - Challenges and Opportunities}
		The presented efforts gather around two focal points: oracle-less and oracle-guided attacks. The intrinsic mechanisms of these attacks shed light on major vulnerabilities in existing logic locking that are exploitable by ML. We summarize the observations as follows:
		
		\subsubsection{Structural vs. Functional Analysis}Oracle-guided attacks focus on functional aspects of schemes, whereas structural leakage is exploited in the oracle-less model due to the absence of an activated IC. This indicates two pitfalls. First, the evaluated schemes have a predictable effect on the functionality of the circuit for incorrect keys, enabling the possibility to perform a guided heuristic search of the key space. Second, the existing schemes induce structural changes which strongly correlate with the value of the key. A mitigation depends on what an attempted attack tries to exploit. For example, to overcome SAIL or SnapShot, a scheme must not reveal anything about the correctness of the key through the induced structural change. This is achieved if the inserted change does not differ depending on the key value. Similarly, to protect against GNNUnlock, the deployed schemes have to ensure resilience against isolation, i.e., the locking circuitry must not be structurally independent from the original design. In the case of GenUnlock and the PSO attack, the behavior of the underlying scheme must be constant or fully random for all incorrect key values; disabling any chance for a guided heuristic convergence towards a correct key. Similar observations can be made for the other attacks as well. Nevertheless, conveying all necessary security objectives into a uniform locking scheme remains an open challenge.
		
		
		\subsubsection{Logic Synthesis for Security}The reliance on logic synthesis transformations to enable the security of logic locking schemes has to be revised. As shown in SAIL, the synthesis rules are predictable and reversible, thereby having little impact on deducing a correct key for traditional schemes. Nevertheless, resynthesis can increase the difficulty to reverse engineering the functionality of a design. However, this needs further evaluation in the context of novel locking policies.
		
		\subsubsection{Structural Isolation of Post-SAT Schemes}PSLL schemes induce isolable changes which create a clear structural distinction between the locking-related gates and the original (regular) gates. The main reason for this vulnerability is the requirement of achieving SAT-resistant behavior for incorrect keys. This specific functional pattern requires significant structural netlist changes consolidated in isolable components (e.g., the tree of complementary logic in Anti-SAT or the restore/perturb units in TTLock and SFLL-HD).
		
		\subsubsection{Output Uncertainty}Regardless of the attack type, ML-based attacks tend to generate an approximate output (as marked in Table~\ref{table:attacks}), meaning that the retrieved key or deobfuscated circuit cannot be evaluated as correct with an exact certainty. Note that this is not the case for NNgSAT. Nevertheless, the approximate output can be utilized as seed for other attacks to speed up the reverse engineering procedure.
		
		\subsubsection{RTL vs Gate-Level}As shown in Table~\ref{table:attacks}, the existing attacks focus on extrapolating information leakage from gate-level netlists. So far, it is unclear if the same leakage is present in RTL-based locking schemes as well.
		
		\subsubsection{Overspecialization}Available mitigation schemes suffer from overspecialization for thwarting specific attack vectors. So far, it has not been evaluated which form of logic locking offers a comprehensive protection against any form of ML-based guessing attack. Hence, a potentially fruitful opportunity lies within the automatic design of resilient schemes~\cite{LeGO2021}.
		
		\subsubsection{ML for Sequential Locking}To the best of our knowledge, ML-based techniques have not been deployed yet for the evaluation of sequential logic locking techniques.

		\subsubsection{The Need for Benchmarks} There is still a need to have access to a rich benchmark set for data-driven analysis~\cite{10.1145/3400302.3415648}.
		
		\section{Conclusion}\label{sec:conclusion}
		This work summarizes the recent developments of logic locking at the frontiers of machine learning. The presented ML-based attacks indicate the presence of structural and functional leakage in contemporary locking schemes which has been overlooked by the traditional interpretation of security. We show that an ML-based analysis is able to pinpoint novel vulnerabilities and challenge existing security assumptions. The offered discussion consolidates the major ML-induced challenges in logic locking design, offering a fruitful ground for future research directions in the post-ML era.

		\bibliographystyle{IEEEtran}
		\bibliography{bibliography_short}

	\end{document}